# Broadband dielectric spectroscopy of Ba(Zr,Ti)O$_3$: dynamics of relaxors and diffuse ferroelectrics

J. Petzelt[1], D. Nuzhnyy[1], M. Savinov[1], V. Bovtun[1], M. Kempa[1], T. Ostapchuk[1], J. Hlinka[1],

G. Canu[2], and V. Buscaglia[2]

[1]Institute of Physics ASCR, Prague, Czech Republic

[2]Institute for Energetics and Interphases, CNR-IENI Genoa, Italy

e-mail petzelt@fzu.cz

**Abstract**

Broadband dielectric spectroscopy from Hz up to the infrared (IR) range and temperature interval 10-300 K was carried out for $x$BaZrO$_3$-(1-$x$)BaTiO$_3$ (BZT-$x$, $x$ = 0.6, 0.7, 0.8) solid solution ceramics and compared with similar studies for $x$ = 0, 0.2, 0.4, 1 ceramics published recently (Phys. Rev. B **86**, 014106 (2012)). Rather complex IR spectra without appreciable mode softening are ascribed to Last-Slater transverse optic (TO) phonon eigenvector mixing and possible two-mode mixed crystal behavior. Fitting of the complete spectral range requires a relaxation in the 100 GHz range for all the samples. Below 1 GHz another relaxation appears, which is thermally activated and obeys the same Arrhenius behavior for all the relaxor BZT samples. The frequently reported Vogel-Fulcher behavior in BZT relaxors is shown to be an artifact of the evaluation from the permittivity or loss vs. temperature dependences instead of its evaluation from loss vs. frequency maxima. The relaxation is assigned to local hopping of the off-centered Ti$^{4+}$ ions in the frozen BTO clusters, whose size is rather small and cannot grow on cooling. Therefore BZT is to be considered as a dipolar glass rather than relaxor ferroelectric.

**Short title:** Broadband dielectric spectroscopy of Ba(Zr,Ti)O$_3$

**Introduction**

$x$BaZrO$_3$-(1-$x$)BaTiO$_3$ (BZO-BTO) solid solutions (BZT-$x$) have become recently extremely popular as one of the possible lead-free piezoelectric and relaxor ferroelectric systems with a potential use for applications as microwave (MW) electric-field tunable materials [1,2] and from the fundamental physics point of view as a prototype isovalent mixed system with relaxor ferroelectric properties. The BZT-x ceramics can be prepared for all $0 \leq x \leq 1$ and by increasing $x$ from $x = 0$ (neat barium titanate BTO) it passes from conventional ferroelectric behavior similar to BTO with a decreasing ferroelectric transition from $T_C \approx 405$ K to ~350 K for $x = 0.15$, via diffuse ferroelectric transition from about $x = 0.15$ to ~0.25 ($T_C \approx 350$-280 K) to a clear relaxor behavior for $x = 0.25$ up to ~0.75 (maximum permittivity temperature $T_m \approx 280$-130 K) and dipolar glass behavior for higher $x$ up to 0.95 ($T_m \approx 130$-80 K)[3,4] and finally weak incipient ferroelectric (paraelectric) behavior for neat BaZrO$_3$ (BZO) [5]. Presently, it is the best studied isovalent solid solution with relaxor behavior, which attracted also a number of recent first-principles studies [6-10]. Extended X-ray absorption fine structure (EXAFS) spectroscopy and neutron total scattering studies in combination with density functional theory have revealed [6-8,11] that the local Zr-O and Ti-O distances are essentially independent of the composition $x$, which is supported by the linear unit-cell volume increase from 64.5 Å$^3$ for $x = 0$ to 73.8 Å$^3$ for $x = 1$, due to a larger ionic radius of Zr$^{4+}$ (0.72 Å) compared to Ti$^{4+}$ (0.605 Å). This results in off-centering of the Ti$^{4+}$ ions in contrast to centric Zr$^{4+}$ positions within the BO$_6$ octahedra for the whole composition range. No segregation of Zr or Ti atoms was detected for the intermediate $x$ concentrations. Their distribution is naturally frozen at usually studied temperatures [10], certainly below the Burns temperature of ~450 K [2,3]. This means that the polar nanoregions (PNR) in BZT, if any, are fixed to the frozen distribution of local BTO unit cells and cannot increase above the size of small BTO (chemical) clusters. This essentially differs from the Pb-containing relaxors, where the PNR (consiting mainly of correlated off-centered Pb positions) are more or less independent of the chemical nanoclusters, which consist of ordered B'-B'' atoms regions [12] and therefore their microscopic picture is more complex and still disputed [13].

Dielectric spectroscopy of the BZT ceramic system was thoroughly investigated using standard capacitance/impedance measurements technique in the frequency range below ~1 MHz (hereafter low-frequency (LF) range) and several times published, see e.g. [1-4] and some additional recent publications refered to in [14]. At higher frequencies the published data have been much more scarce, but recently we reported a detailed study on rather broad dielectric response from the Hz range up to the infrared (IR) and temperatures 10-700 K for the diffuse ferroelectric BZT-0.2 and relaxor BZT-0.4 ceramics, compared with similar data on endmember BTO and BZO ceramics [14]. BZO behaves as a rather standard cubic perovskite dielectric with three sharp polar phonon modes (expected from the symmetry) and no dielectric dispersion at lower frequencies. The weak incipient ferroelectric behavior (increase in the permittivity from ~36 at 300 K to ~40 at 10 K) is due to a small softening of the transverse optic TO1 mode from 121 cm$^{-1}$ at 300 K to 116 cm$^{-1}$ at 10 K. However, the strongest mode (mode with the highest oscillator strength or mode plasma frequency), known as the Slater mode [15], is the second TO2 mode in agreement with first-principles calculations [5], unlike in BTO, where it is the lowest frequency TO1 mode (soft mode). Therefore the phonon eigenvector mixing is expected for the BZT solid solution, which has manifested itself in the IR spectra as a rather complex IR response with several, only weakly temperature dependent, additional peaks in the phonon response, particularly in the TO2 range. No appreciable softening was observed towards the dielectric maximum temperature $T_m$ or $T_C$. Moreover, additional complex broad dielectric dispersion appeared in BZT below the THz range with the high-frequency wings apparent from our time-domain THz transmission data, reminding the central mode (CM) phenomena previously observed in BTO single crystals [16, 17], which we have confirmed also for BTO ceramics in all the phases (including ferroelectric ones). These features gradually weaken on cooling vanishing towards the lowest temperature.

Analysis of the dielectric dispersion below the THz range indicated one or two broad Cole-Cole relaxations in the dielectric spectra above the MHz range. Since we did not have experimental data between about 1 and 150 GHz, we could not accurately determine the dispersion in this range. However, the joint fitting of the real and imaginary part of the dielectric response from the low-frequency Hz range up to 2 THz required rather two peaks above 1 GHz in the loss spectra. The lower-frequency peak slows down on cooling, whereas the higher-frequency one only weakens, but stays in the 100 GHz range.

The dispersion below 1 GHz essentially differs for both BZT samples. In the diffuse ferroelectric BZT-0.2 the GHz peak merges into a nearly constant loss background below the diffuse ferroelectric phase transition at $T_C \approx 300$ K, whose level gradually decreases on cooling. This explains the basic feature of the diffuse transition that the permittivity vs temperature maximum at $T_C$ remains frequency independent, even if the value of permittivity maximum decreases with increasing frequency. We observed that this was valid up to ~1 GHz only. In the GHz range one could expect similar shift of the $T_m$ as in usual relaxors, due to the slowing down of the GHz relaxation. The nearly constant loss level below $T_C$ is then interpreted as the dynamics of ferroelectric nanodomains, more precisely dynamics of nanodomain boundaries (breathing of nanodomains), whose broadness should be assigned to broad distribution of nanodomain sizes, shapes and pinning.

In the relaxor BZT-0.4 sample the lower-frequency relaxation slows down from ~450 K (Burns temperature), where it emerges from the GHz dispersion, continues softening even below $T_m \approx 150$ K and shows thermally activated behavior, accompanied with its broadening and finally also merging into a constant loss backgorund (lower than for BZT-0.2) below ~100 K. Careful fitting of its mean frequency ν of the maxima from ε''(ν) spectra shows a good Arrhenius fit $\nu = \nu_0 \exp[-U/T]$ with activation energy $U = 0.167$ eV and attempt frequency $\nu_0 \approx 1$ THz. Similar fits using maxima from the ε'(T) and ε''(T) dependences yield rather Vogel-Fulcher law $\nu = \nu_0 \exp[-U/(T-T_{VF})]$ with the Vogel-Fulcher temperatures $T_{VF}$ = 45.6 and 98.1 K, respectively, and the same $\nu_0$. This clearly manifests the well-known [18, 19], but often not considered fact, that $T_{VF}$ is in fact frequently higher than the actual freezing temperature. This can be easily understood due to the fact that the dielectric strength of the main relaxation is decreasing on cooling. Since the main low-frequency dielectric relaxation in BZT should be due to the dynamics of the off-centered Ti ions in the BTO cells, the simple thermal activation of the low-frequency relaxation indicates a simple non-correlated hopping of them. This reminds more dipolar glass behavior than the classical relaxor behavior. To check how such a behavior develops with BZT composition, we have prepared additional ceramics BZT-0.6, 0.7 and 0.8 and investigated their broad dielectric response in similar way as in Ref. [14]. To enable a better polar phonon assignment, we shall also discuss and assign the polar phonon modes in all the phases of BTO ceramics, as evaluated from our previous measurements [14, 20].

**Experiment**

BZT-0.6, 0.7, 0.8 powders were prepared using the standard solid state reaction by calcination at 1000°C (4 hours) and the ceramics were processed by cold isostatic pressing (1500 bar) and sintering at 1600°C for 4 hours. The ceramics were of single phase perovskite structure with over 99% theor. density and grain size ~1 μm. Their scanning electron microscope (SEM) pictures on fractured surfaces are shown in Fig. 1.

Dielectric response was studied between 10 and 300 K in the frequency range of 1 Hz – 90 THz. Five experimental techniques were used to cover such a broad frequency range. Two techniques were used for samples with sputtered Au electrodes: standard LF capacitance measurements in the $1$–$10^6$ Hz range using frequency analyzer Novocontrol Alpha AN and dielectric response of cylindrical BZT samples (~0.8 mm diameter and ~5 mm length) in the HF range (1 MHz–1 GHz) using dielectric spectrometer with Agilent 4291B impedance analyzer and Novocontrol BDS 2100 coaxial sample cell and Sigma System M18 temperature chamber. Three electrode-less techniques were used in the MW-THz-IR range: Composite dielectric resonator technique in the 5-10 GHz range using vector network analyzer Agilent E8364B, time-domain THz transmission spectroscopy (0.15-2 THz) using home-made THz spectrometer based on Ti:sapphire femtosecond laser with polished thin BZT pellets of ~110 μm (BZT-0.8) and ~56 μm thickness (BZT-0.6, 0.7), and IR reflectivity spectroscopy on thicker polished pellets using the Fourier-transform infrared (FTIR) spectrometer Bruker IFS 113v (1-20 THz). An Optistat continuous-flow cryostat with thin mylar and thick polyethelene windows was used for cryogenic THz and IR measurements, respectively.

**Results and discussion**

In Fig. 2 the IR reflectivity spectra of the new BZT ceramics at 300 and 10 K are compared with those measured in Ref. [14], including the THz data calculated from the THz transmission measurements and the joint fits with standard generalized oscillator model using factorized form of the dielectric function [14]. The TO frequencies for the mixed BZT ceramics, as evaluated from our fits above ~50 cm$^{-1}$, are shown in Fig. 3. One can see that the temperature changes (except for the THz range) are not very pronounced. No pronounced temperature dependences and no softening of any mode, as already discussed in Ref. [14], can be observed. This is particularly surprising for BZT-0.2 with the diffuse transition into the rhombohedral phase, which should be therefore considered as predominantly of the order-disorder type [14]. The behavior at THz and lower frequencies will be discussed below.

The absence of softening and temperature independence of phonons strongly differ from the behavior in BTO with the three ferroelectric phases. In Fig. 4 we present the plot of all pronounced TO phonon peak frequencies in BTO for all phases from 850 K down to 10 K, together with their suggested assignment. One can see the huge $A_1$–E splitting of the TO1 soft mode at the para-ferro transition $T_C$, known also from single crystals [17, 21], and appearance of the CM below the soft mode response already more than 400 K above $T_C$, as confirmed by first principles calculations [22], which remains detectable in our THz spectra down to the transition to rhombohedral phase. This CM softens as expected near $T_C$ and is essentially responsible for the dielectric Curie-Weiss anomaly. The plotted frequencies in Fig. 4 below 100 cm$^{-1}$, where the modes are heavily damped or overdamped, correspond to dielectric loss peaks rather than to fitted mode frequencies $\nu_i$, which are always somewhat higher, but its determination is rather inaccurate (similar fits can be achieved by varying $\nu_i$ and $\gamma_i$ keeping the ratio $\nu_i^2/\gamma_i$ constant, where $\gamma_i$ is the damping of the i-th mode). The existence of the low-frequency and low-temperature CM in the orthorhombic and rhombohedral phases (denoted in Fig. 4 by empty circles) is suggested from the low-frequency loss wings in our THz spectra, which are seen down to the lowest temperature.

Concerning the behavior in the ferroelectric phases, we note that for anisotropic grains the evaluation of phonon frequencies is not quite straightforward since all the principal dielectric responses are intermixed and maybe influenced by depolarizing fields on the grain boundaries. But our experience with other ceramics, e.g. $Ba_{0.7}Sr_{0.3}TiO_3$ [23] shows that the TO frequencies are almost not influenced, unlike their oscillator strengths and LO frequencies, which are strongly influenced and cannot be estimated at all from the spectra of ceramics by a simple fitting. Only modeling using effective medium approach (EMA) may attempt to estimate separately the dielectric functions along the principal axes directions in ceramics with anisotropic grains or clusters [23, 24]. Such a detailed comparison of our IR-THz spectra of BTO ceramics with those from single crystals is beyond the scope of the present study. Also a more detailed discussion of the phonon behavior in the BZT mixed system, where the mixing of eigenvectors between the TO1 and TO2 modes occurs combined with two mode behavior in the TO2 mode region [14] is postponed for another study.

In Fig. 5a, b we present our dielectric spectra of BZT-0.6, 0.7 and 0.8 for 3 selected temperatures in a broad frequency range $10^3$-$10^{14}$ Hz, together with their joint fit for the real and imaginary part of the dielectric function (to check the Kramers-Kronig relations). The fit necessarily requires a presence of a pronounced loss peak in the 100 GHz range for all samples, even if direct measurements in this range are not available. This peak, which weakens on cooling, was fitted with a simple Debye relaxation with temperature independent frequency. Similar peak was also needed for fitting the BZT-0.2 and 0.4 [14]. If only one loss peak would be assumed in the whole GHz-THz range, the low-frequency permittivity would be much higher than actually measured.

The second, lower frequency relaxation peak appears in the GHz range at room temperature. On cooling it strongly softens and broadens and finally vanishes in the small constant background losses at low temepratures. For all three BZT samples it was fitted with a Cole-Cole relaxation $\varepsilon(\omega) = \dfrac{\Delta\varepsilon}{1+(i\omega\tau)^{1-\alpha}}$ with α increasing on cooling, which describes the broadening of the loss peak. Its relaxation frequency $\nu \equiv 1/(2\pi\tau)$ is thermally activated, obeying the Arrhenius law with the same activation energy $U \approx 0.16$ +/-0.01 eV and limiting $\nu_0 \approx 1$ THz for all compositions, as found also for the BZT-0.4 sample [14]. Some weaker secondary relaxations, also thermally activated, but with parameters slightly varying for different samples ($U \approx$ 0.05-0.13 eV), can be also detected in our spectra. All the observed relaxation frequencies for all our BZT compositions are plotted in Fig. 6.

Let us now discuss the obtained results on relaxations. The most striking feature is the indirectly seen relaxation in the 100 GHz (~3 cm$^{-1}$) range, to our knowledge observed for the first time in relaxors, since no reliable dielectric data are available for this frequency range for any relaxor ferroelectric. Its detection requires at least a very broad dielectric spectrum around this range, which has not been either available for other systems. Also the THz loss wing in pure BTO ceramics at low temperatures (Fig. 4) could be due to a similar relaxation. We assign it to so-called quasi-Debye loss mechanism, theoretically predicted for noncentrosymmetric structures as due to fluctuation of the thermal phonon density distribution function [25, 26]. Its frequency is roughly given by the mean damping of thermal phonons. Its dielectric activation in our samples could be ascribed to local acentricity of BTO.

The thermally activated relaxation, which follows a common Arrhenius law for all the BZT samples of the relaxor composition (0.4 ≤ x ≤ 0.8), is the most important feature from our studies. It shows that the previously evaluated Vogel-Fulcher law [1-4] and some other recent papers cited in Ref. [14] is an artifact of evaluation from the $\varepsilon'(T)$ or $\varepsilon''(T)$ dependences rather than using the proper evaluation from the $\varepsilon''(\omega)$ spectra as a function of T, which, however, requires a broader frequency range. Since this relaxation in BZT can be safely assigned to hopping of the off-centered Ti$^{4+}$ ions (see also the first principles calculations [10]), its simple thermal activation indicates an uncorrelated hopping of individual off-centered Ti ions around their centrosymmetric sites over a temperature independent energy barrier of 0.16 eV. In this way, BZT can be classified as a simple dipolar glass rather than relaxor ferroelectric. This picture agrees with the impossibility to induce a stable polarization in a field-cooled regime (at 40 kV) for BZT-x compositions with x ≥ 0.5 [3]. The weaker secondary relaxations, also shown in Fig. 6, are also thermally activated, but their frequencies differ for different compositions. We assign them temporarily to Ti-hopping, as well, but in the boundary regions between BTO and BZO clusters, which may depend on the specific geometry of these clusters in the samples.

## Conclusions

The broad-band dielectric spectroscopy on BZT-x ceramics for a rather complete composition range has revealed several surprising important features. Unlike in neat BTO, where the softening of THz CM accounts for the dielectric Curie-Weiss anomaly, no THz-IR mode softening was detected for any composition of the solid solution. Below the THz range a relaxation with temperature independent frequency of ~100 GHz was suggested from our fitting procedure and could be assigned to quasi-Debye loss mechanism, induced by local broken centrosymmetricity. Dielectric maxima in the temperature dependences are due to slowing down of another relaxation, which appears near the GHz range at room temeprature, slows down on cooling and is thermally activated obeying a simple Arrhenius law with the same activation energy ~0.16 eV for all relaxor compositions. We assign it to a simple local hopping of the off-centered $Ti^{4+}$ ions around their centrosymmetric positions in the frozen BTO clusters. The polar nanoregions in BZT are therefore identical with the clusters of BTO in the BZO matrix, which are known to be rather small (BZT forms rather homogeneous solid solution), remain frozen from very high temperatures and cannot grow on cooling. So, BZT shows clearly properties of a dipolar glass rather than of a relaxor ferroelectric.

## Acknowledgements

This research was supported by the Czech Science Foundation (projects Nos. P204/12/0232 and 13-15110S).

**Figure caption**

Fig. 1 SEM pictures of BZT-0.6, 0.7, and 0.8 ceramics on the fracture surfaces.

Fig. 2 IR reflectivities of the whole BZT ceramic system at room temperature and 10 K. Calculated THz data (full symbols) and fits (solid lines) are included.

Fig. 3 Temperature dependence of the polar phonon peak frequencies of BZT evaluated from our IR-THz fits.

Fig. 4 Temperature dependence and assignment of the TO polar phonon frequencies in BTO ceramics.

Fig. 5 Dielectric functions of BZT-0.6, 0.7, 0.8 in a broad frequency range and their fits at selected temperatures (a) real part - permittivity, (b) imaginary part – losses.

Fig. 6 Temperature dependences of the evaluated relaxation frequencies in BZT ceramics. Empty symbols belong to main (stronger) relaxations, full symbols to secondary (weaker) ones.

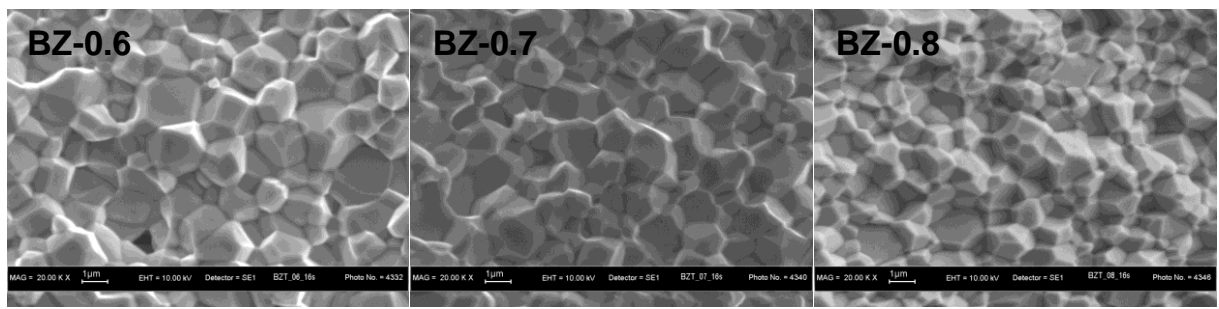

Fig. 1 SEM pictures of BZT-0.6, 0.7, and 0.8 ceramics on the fracture surfaces.

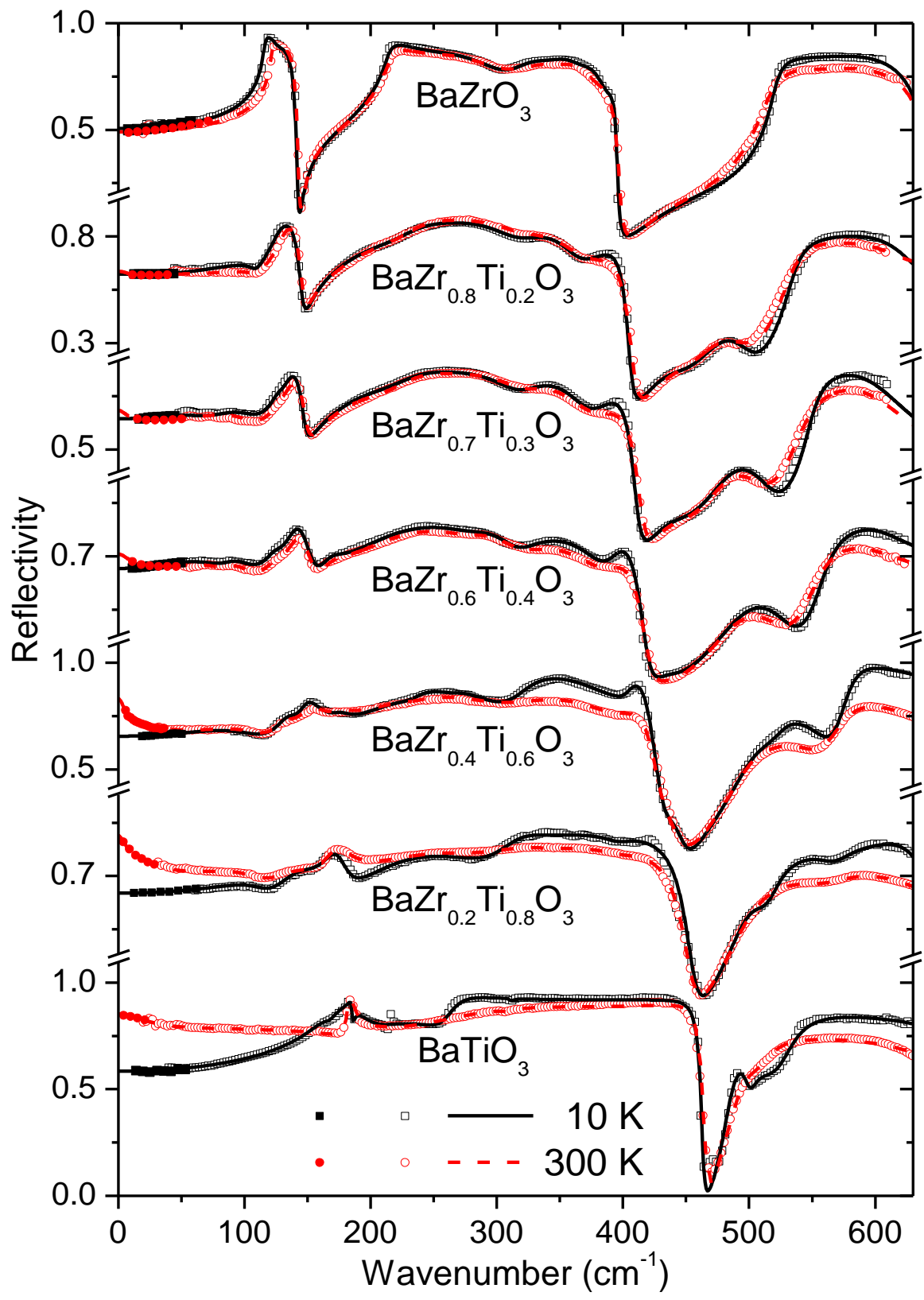

Fig. 2 IR reflectivities of the whole BZT ceramic system at room temperature and 10 K. Calculated THz data (full symbols) and fits (solid lines) are included.

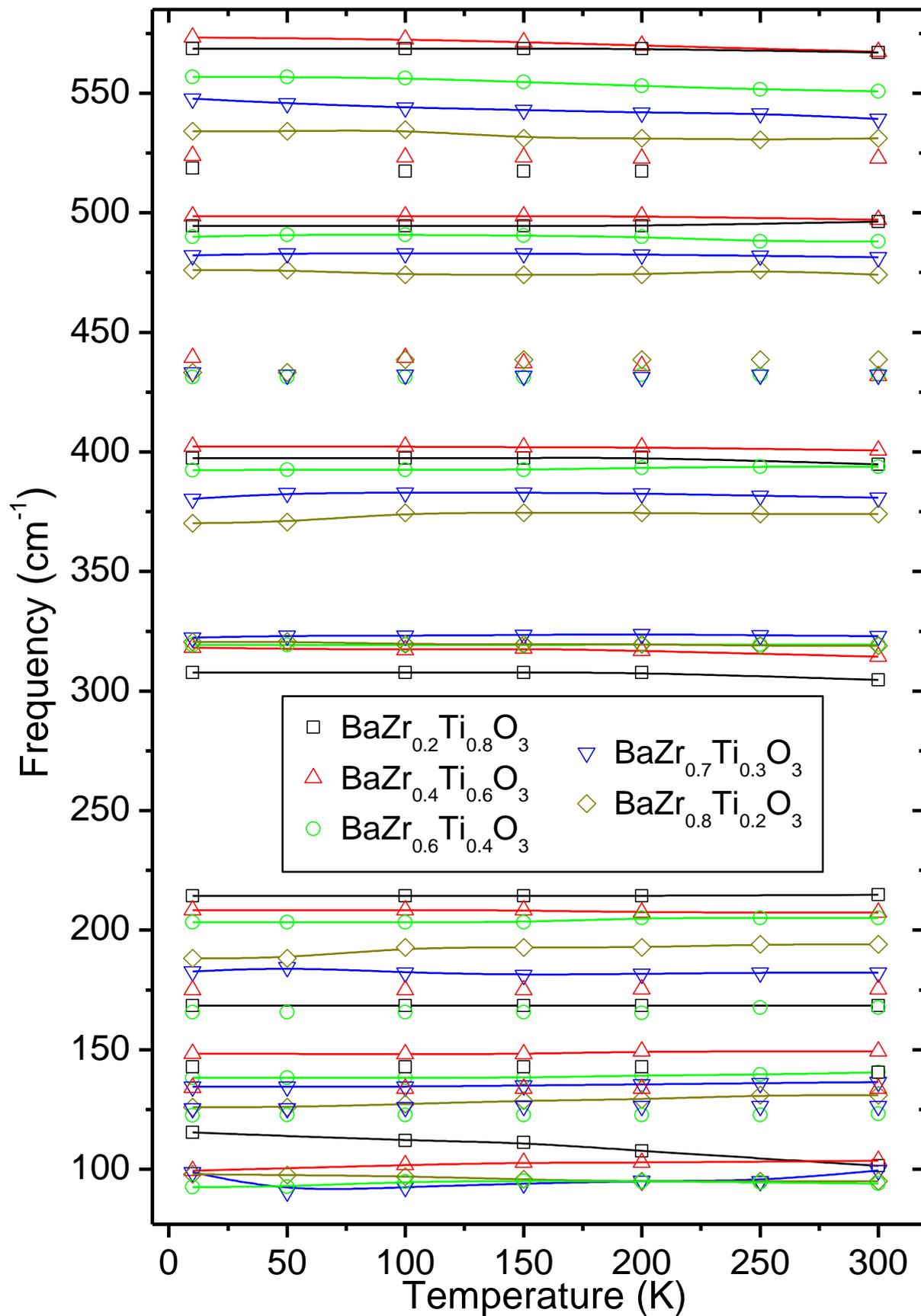

Fig. 3 Temperature dependence of the polar phonon peak frequencies of BZT evaluated from our IR-THz fits.

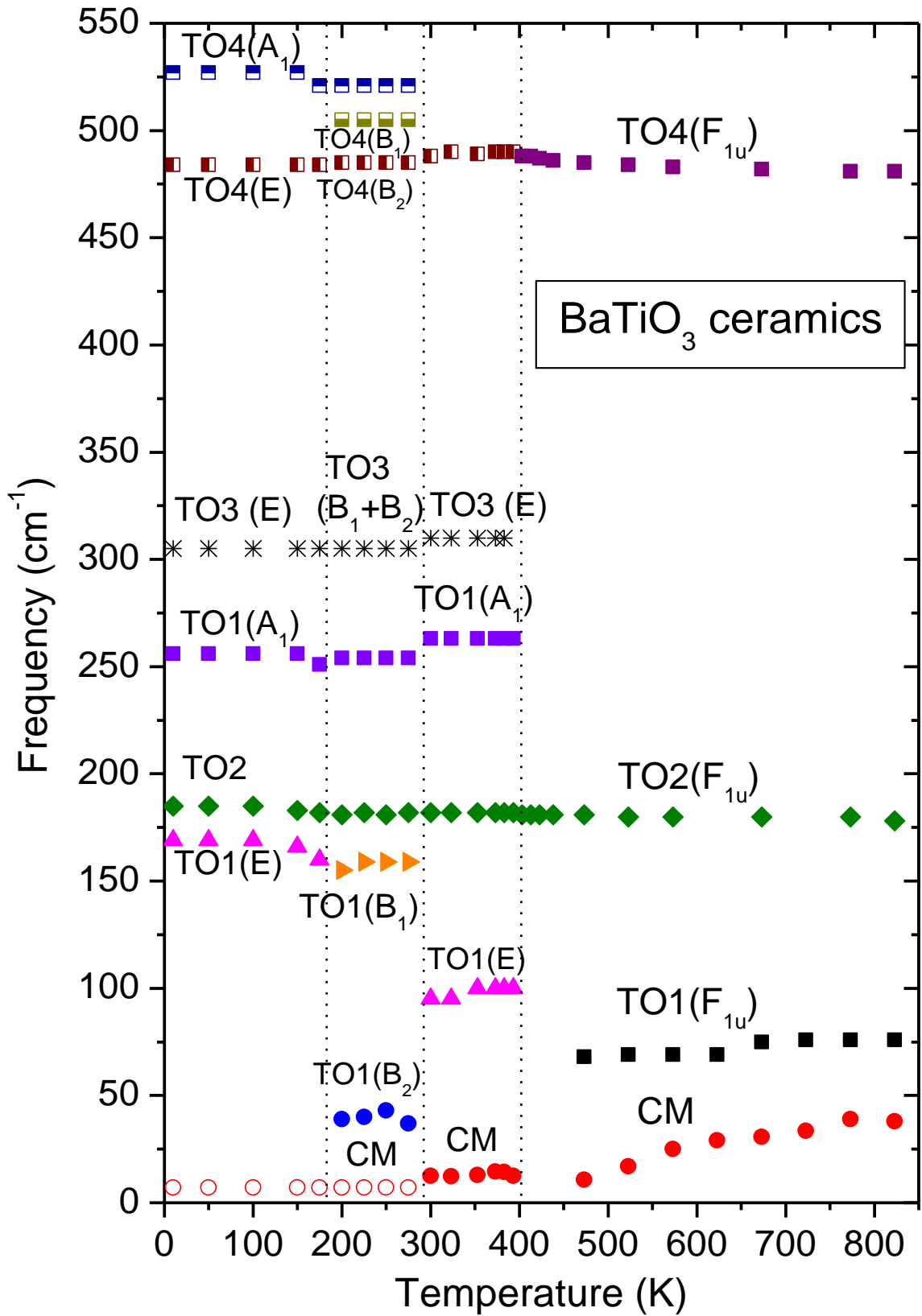

Fig. 4 Temperature dependence and assignment of the TO polar phonon frequencies in BTO ceramics.

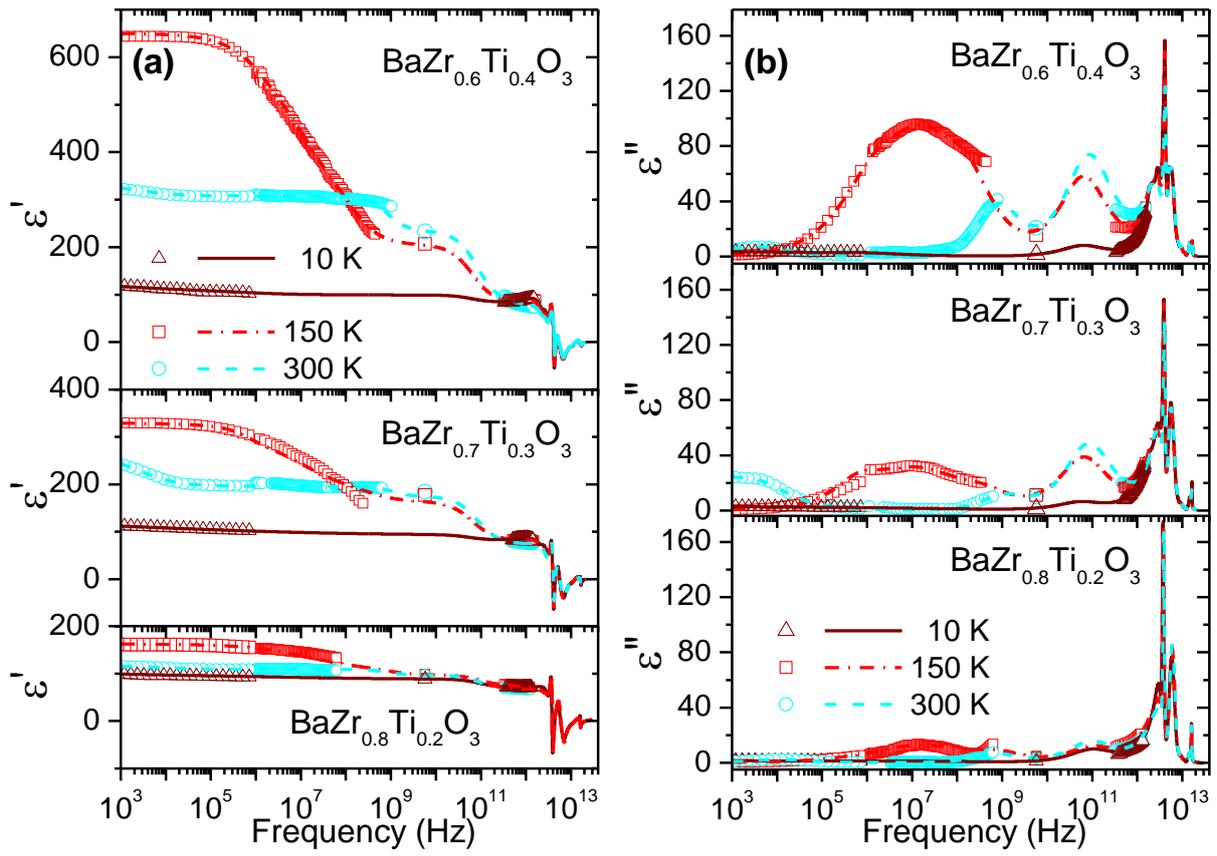

Fig. 5 Dielectric functions of BZT-0.6, 0.7, 0.8 in a broad frequency range and their fits at selected temperatures (a) real part - permittivity, (b) imaginary part – losses.

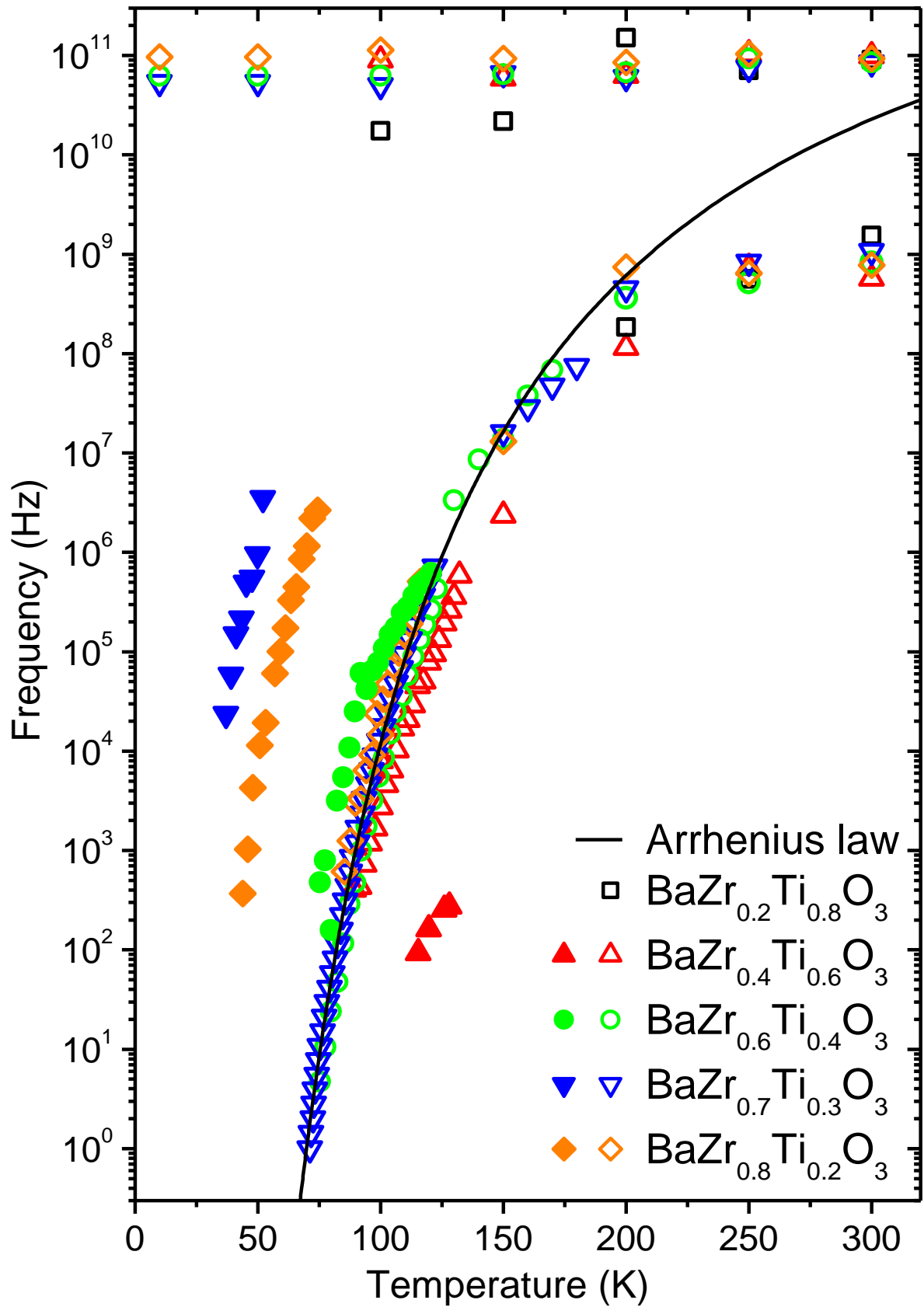

Fig. 6 Temperature dependences of the evaluated relaxation frequencies in BZT ceramics. Empty symbols belong to main (stronger) relaxations, full symbols to secondary (weaker) ones.